\documentstyle[aps,prd,epsf,amssymb]{revtex}

\def \be{\begin{equation}}
\def \ee{\end{equation}}
\def \bea{\begin{eqnarray}}
\def \eea{\end{eqnarray}}
\def \dd{\nabla}
\def \s2{\sqrt 2}

\begin{document}


\tighten

\title{Construction of Astrophysical Initial 
Data for Perturbations of Relativistic Stars}

\author{Nils Andersson${}^{(1)}$, 
        Kostas D. Kokkotas${}^{(2)}$,
        Pablo Laguna${}^{(3)}$,
        Philippos Papadopoulos${}^{(4)}$ and
        Michael S. Sipior${}^{(3)}$}

\address{
${}^{(1)}$ Department of Mathematics\\
University of Southampton, Southampton SO17 1BJ, UK }

\address{
${}^{(2)}$ Department of Physics\\
Aristotle University of Thessaloniki\\
Thessaloniki 54006, Greece}

\address{
${}^{(3)}$ Department of Astronomy \& Astrophysics and \\
Center for Gravitational Physics \& Geometry\\
Penn State University, University Park, PA 16802, USA}

\address{
${}^{(4)}$ Max-Planck-Institut f\"ur Gravitationsphysik\\
Am Muhlenberg 5, 14476 Golm, Germany}

\maketitle

\begin{abstract}

We develop a framework for constructing initial data sets for
perturbations about spherically symmetric matter distributions.  
This framework facilitates setting initial data representing
astrophysical sources of gravitational radiation involving
relativistic stars.  The procedure is based on Lichnerowicz-York's conformal
approach to solve the constraints in Einstein's equations.  The
correspondence of these initial data sets in terms of the standard
gauge perturbation variables in the Regge-Wheeler perturbation variables is
established, and examples of initial data sets of merging
neutron stars under the close-limit approximation are presented.

\end{abstract}

\widetext

\section{Introduction}
\label{sec:introduction}

The early years of the next millennium will hopefully be remembered 
for the birth of Gravitational-wave Astronomy. 
With several large scale interferometers (LIGO, VIRGO, GEO600, TAMA)
under construction, and the continued improvement of the technology for 
cryogenic resonant-mass detectors (ALLEGRO, AURIGA, EXPLORER, NAUTILUS), 
there are many reasons to be optimistic at the present
time.  However, the interpretation of
data from the new generation of detectors 
will heavily depend on accurate ``templates'' of
gravitational wave-forms.  For a given astrophysical source of
gravitational radiation, construction of such templates involves
fully non-linear or perturbative approximations to Einstein's field
equations.  In both instances, the construction of appropriate initial data
constitutes a fundamental issue. It is absolutely necessary
that initial data represent a ``realistic'' stage of the astrophysical
system under consideration. 

The early years of Numerical Relativity were in part characterized by
studies aimed at constructing initial data for Einstein's equations,
namely data that satisfies the Hamiltonian and momentum constraints.
Of particular interest was obtaining solutions to the constraints which
represent black hole binaries \cite{cook}.  These initial data studies
highlighted the importance that Lichnerowicz-York's conformal approach
\cite{york79} plays in facilitating solving the constraints.

In the perturbative arena, initial data sets also play a fundamental
role if a connection with systems of astrophysical relevance is to be
made.  Examples of perturbative studies where astrophysically
consistent initial data is needed are the point-particle \cite{particle}
and close-limit approximations \cite{price94} to black hole
coalescences. For the close-limit approximation in particular, this
issue is crucial. The focus is then on the late 
stage of the merger, when the binary
system can be approximated as a single, perturbed black hole. In the case of 
perturbations of relativistic stars, 
most of the studies 
\cite{allen} have not considered initial data with direct
connection to a given astrophysical situation. In other words, the focus
has so far been on investigating how the star reacts to generic
perturbations.

The goal of this paper is to provide a mechanism for generating
initial data for perturbations of 
a relativistic star. This data should ideally represent 
astrophysical situations of relevance to gravitational-wave
detectors.  Inspired by the success that Lichnerowicz-York's conformal approach
has enjoyed for solving Einstein's constraints, we base our methodology
on ``linearizing Lichnerowicz-York's procedure''. By doing so, we take advantage of
the prescription for knowing which pieces of information, among the
metric and its ``velocity'', are fixed by the constraints and which
are freely specifiable. Moreover, we inherit the machinery used
in the past for the construction of initial data sets representing
binary systems.  That is, the procedure described in this paper
provides a natural framework for obtaining initial data in connection with
the close-limit approximation to neutron star mergers
\cite{nsmergers}.  As with black hole binaries, the close-limit to
neutron star collisions deals with the late stages of the merger, at
the point in which the systems can be approximated as a single neutron
star ``dressed'' with perturbations.

\section{Linearization of Lichnerowicz-York's Conformal Approach}
\label{sec:linearization}

Given a spacetime with 4-metric $\hat g_{\mu\nu}$ and a foliation of
this spacetime with Eulerian observers having a 4-velocity $\hat
n^\mu$, the constraints in Einstein's fields equations can be written
in a 3+1 (or ADM) form \cite{adm} as 
\bea \hat R + \hat K^2 - \hat K_{ij} \hat
K^{ij} & = 16\, \pi\, \hat \sigma\hspace{0.2 true in}
&\hbox{(Hamiltonian)}
\label{eq:ham} \\
\widehat \nabla_{j}(\hat K^{ij} - \hat h^{ij} \hat K) & 
= 8\,\pi \hat j^{i}.\hspace{0.2 true in} &\hbox{(Momentum)}
\label{eq:mom}
\eea 
Above, $\hat h_{ij}$ is the 3-metric ($\hat h_{\mu\nu} = \hat
g_{\mu\nu} + \hat n_\mu\,\hat n_\nu$) and $\hat K_{ij}$ the extrinsic
curvature of the time-like hypersurfaces in the foliation, with $\hat
K = \hat h^{ij}\hat K_{ij}$ and $\hat R$ the scalar curvature.
Furthermore, $\widehat\nabla_{i}$ is the covariant derivative
associated with the 3-metric $\hat h_{ij}$, and $\hat \sigma$ and
$\hat j^i$ are the energy and momentum densities of the matter
sources. Greek letters denote spacetime indexes and Latin letters
spatial indices, and we use units in which $G = c = 1$.  Contrary to the
common practice, we use ``hats'' to denote physical space since
we will be mostly working in the conformal space.

For a perfect fluid, the stress-energy tensor is given by
\be
\hat T_{\mu\nu} = (\hat\rho+\hat p)\,\hat u_\mu\,\hat u_\nu
+ \hat p\,\hat g_{\mu\nu},
\ee
where $\hat\rho$ and $\hat p$ are respectively the 
total mass-energy density and pressure of the fluid measure by an
observer with 4-velocity $\hat u^\mu$.
The energy and momentum densities appearing in the constraints are obtained
from
\bea
\hat\sigma & = \hat T_{\mu\nu}\,\hat n^\mu\,\hat n^\nu 
& = (\hat \rho + \hat p)\, \hat\gamma^2 - \hat p
\label{eq:sigma}\\
\hat j^\mu & = -\hat T_{\nu\alpha} \,\hat n^\nu \, \hat  h^{\alpha\mu}
& = (\hat \rho + \hat p)\,\hat\gamma\,\hat  h^{\mu\nu}\hat u_\nu
\label{eq:j}
\eea
where $\hat\gamma \equiv -\hat n_\mu\,\hat u^\mu$ is the relativistic
boost factor.

The fundamental virtue of Lichnerowicz-York's conformal approach \cite{york79}
for solving the
Hamiltonian and momentum constraints is that it provides a
concrete recipe for singling out which four ``pieces'' among
the twelve components $(\hat h_{ij},\,\hat K_{ij}$) are to be solved
from Eqs.~(\ref{eq:ham}) and (\ref{eq:mom}).
The starting point is the Ansatz 
\be
\hat h_{ij} = \phi^4 h_{ij},
\label{eq:anzats}
\ee
where the conformal metric $h_{ij}$ is assumed to be
known. Thus, for the metric, the piece that is fixed by the
constraints (Hamiltonian) is the conformal factor $\phi$.
The other three quantities fixed by the constraints (momentum)
involve the extrinsic curvature. The idea here is to 
decompose the extrinsic curvature into its trace, tracefree-transverse
and tracefree-longitudinal parts. To achieve this,
the extrinsic curvature is first split into 
\be
\hat K^{ij} = \hat A^{ij} + \frac{1}{3} \hat h^{ij} \hat K.
\label{eq:kij}
\ee
Before decomposing the tracefree part $\hat A^{ij}$ into its transverse and longitudinal
parts, the following conformal transformation is applied:
\be
\hat A^{ij} = \phi^{-10} A^{ij}.
\label{eq:aij}
\ee 
The exponent in the conformal transformation is motivated by the
fact that this transformation possesses the following property: 
\be
\widehat\nabla_j\hat A^{ij} = \phi^{-10} \nabla_j A^{ij},  
\ee 
with $\dd_{i}$ covariant differentiation associated with the
background metric $h_{ij}$. This property simplifies the conformal
transformation of the divergence of the extrinsic curvature in the
momentum constraint (\ref{eq:mom}).  

Since, as will become clear below, the trace of the extrinsic
curvature $\hat K$ is not fixed by the constraints, no conformal
transformation is imposed on the trace of the extrinsic curvature
($\hat K = K$).  Once the conformal transformation is applied, the
next step is to decompose $A_{ij}$ into its transverse and
longitudinal parts, namely
\be
A^{ij} = A^{ij}_{*} + (\l\,W)^{ij},
\ee
where
\bea
(\l\,W)^{ij} & = & 2\,\nabla^{(i}W^{j)} - \frac{2}{3} h^{ij} \dd_{k} W^{k}, \\
\nabla_j A^{ij}_* & = & 0.
\eea
With the above conformal transformations and transverse-longitudinal
decompositions, the Hamiltonian and momentum constraints become
\bea
8 \dd^i \dd_{i} \phi - R\, \phi + A_{ij} A^{ij}
\phi^{-7} - \frac{2}{3} K^2 \phi^5 + 16\,\pi \sigma\, \phi^{-3} & = & 0 
\label{eq:hamc}\\
\dd^j \dd_{j} W^{i} + \frac{1}{3} \dd^i \dd_j W^j + R^{i}{}_{j} W^j
- \frac{2}{3} \phi^6\, \dd^{i} K  - 8\,\pi\, j^{i}&  =&  0\,,
\label{eq:momc}
\eea
where $R^{i}{}_{j}$ is the 3-Ricci tensor of the conformal
space and $R$ its trace. In deriving Eqs.~(\ref{eq:hamc}) and (\ref{eq:momc}), the following
conformal transformations
for the energy and momentum densities were used: 
\bea
\hat\sigma & = & \phi^{-8}\,\sigma 
\label{eq:csigma}\\
\hat j^i & = & \phi^{-10}\, j^i.
\label{eq:cj}
\eea

To summarize, the Hamiltonian constraint fixes the conformal factor $\phi$
and the momentum constraint determines the generator $W^i$ of the 
longitudinal part of the conformal,
traceless part of the extrinsic curvature. 
The freely specifiable data in this coupled set of equations are
the conformal metric $h_{ij}$, the trace of the
extrinsic curvature $K$, the source functions 
$(\sigma,\, j^i)$, and the divergence free, traceless part
of the extrinsic curvature $A^{ij}_*$, which is hidden in $A_{ij}$ in Eq.~(\ref{eq:hamc}).

Thus far, we have just reviewed Lichnerowicz-York's treatment of the initial data problem.
We now introduce our {\em first} assumption. The initial data,
 $(\hat h_{ij},\,
\hat A^{ij},\, \hat K,\, \hat\sigma,\, \hat j^i)$,
are assumed to be close to a given background, i.e.,
\bea
\hat h_{ij} & = & \hat h_{ij}^{(0)} + \hat h_{ij}^{(1)}\\
\hat A^{ij} & = & \hat A^{ij}_{(0)} + \hat A^{ij}_{(1)}\\
\hat K & = & \hat K_{(0)} + \hat K_{(1)}\\
\hat \sigma & = & \hat \sigma_{(0)} + \hat \sigma_{(1)}\\
\hat j^i & = & \hat j^i_{(0)} + \hat j^i_{(1)},
\eea
where ${}^{(0)}$ labels background quantities and ${}^{(1)}$ first-order
perturbations. The corresponding decomposition of the conformal
space quantities yields:
\bea
h_{ij} & = & h_{ij}^{(0)} + h_{ij}^{(1)} \\
\phi & = & \phi^{(0)} + \phi^{(1)} \\
A^{ij}_{*} & = & A^{ij}_{*(0)} + A^{ij}_{*(1)}\\
W^i & = & W^i_{(0)} + W^i_{(1)} \\
K & = & K^{(0)} + K^{(1)} \\
\sigma & = & \sigma^{(0)} + \sigma^{(1)} \\
j^i & = & j^i_{(0)} + j^i_{(1)},
\eea
where
\bea
\hat h_{ij}^{(0)} & = & \phi^4_{(0)}\, h_{ij}^{(0)}\\
\hat h_{ij}^{(1)} & = & \phi^4_{(0)}\, h_{ij}^{(1)} 
+ 4\,\phi^3_{(0)}\,\phi_{(1)}\,h_{ij}^{(0)}\\
\hat A^{ij}_{(0)} & = & \phi^{-10}_{(0)}\, A^{ij}_{(0)}\\
\hat A^{ij}_{(1)} & = & \phi^{-10}_{(0)}\, A^{ij}_{(1)} 
- 10\,\phi^{-11}_{(0)}\,\phi_{(1)}\,A^{ij}_{(0)}\\
\hat K_{(0)} & = & K_{(0)}\\
\hat K_{(1)} & = & K_{(1)}\\
\hat \sigma_{(0)} & = & \phi^{-8}_{(0)}\,\sigma_{(0)}\\ 
\hat \sigma_{(1)} & = & \phi^{-8}_{(0)}\,\sigma_{(1)}
- 8\,\phi^{-9}_{(0)}\,\phi_{(1)}\,\sigma_{(0)}\\
\hat j^i_{(0)} & = & \phi^{-10}_{(0)}\,j^i_{(0)}\\ 
\hat j^i_{(1)} & = & \phi^{-10}_{(0)}\,j^i_{(1)}
-10\,\phi^{-11}_{(0)}\,\phi_{(1)}\, j^i_{(0)}.
\eea
Although all scalars, vectors and tensors are expanded in order of
smallness, it is important to stress again that all but $\phi$
and $W^i$ are freely specifiable, and this property is independent of
the order of the perturbation.  

At this point we introduce our {\em second} assumption, which is that the
perturbations of the conformal background vanish on the initial
data slice; that is, $h_{ij}^{(1)} = 0$. 
The primary motivation for this choice
is the simplification of the coupled system of constraint equations. 
We shall later discuss the physical
relevance, as well as the implications and restrictions,
that this assumption imposes on the class of initial data
that one has access to with our procedure. Using $h_{ij}^{(1)} = 0$,
together with the above perturbative expansions in the coupled
elliptic system (\ref{eq:hamc}) and (\ref{eq:momc}), we obtain
\bea
8 \dd^j \dd_{j} \phi_{(1)} - \left[
  R  
+ 7 A_{(0)}^{ij} A^{(0)}_{ij}\,\phi_{(0)}^{-8}  
+ \frac{10}{3} K^2_{(0)}\,\phi^{4}_{(0)}
+ 48\,\pi\, \sigma_{(0)} \,\phi^{-4}_{(0)} \right]\phi_{(1)} && \nonumber\\
+\, 2\,A^{(0)}_{ij} A^{ij}_{(1)}\,\phi^{-7}_{(0)}
- \frac{4}{3} K_{(0)} K_{(1)}\,\phi^{5}_{(0)}
- 16\,\pi\, \sigma_{(1)}\,\phi^{-3}_{(0)} & = & 0
\label{eq:haml} \\
\dd^j \dd_{j} W^{i}_{(1)} + \frac{1}{3} \dd^i \dd_j W^j_{(1)} 
+ R^{i}{}_j W^j_{(1)}
- \frac{2}{3} \phi^6_{(0)} \dd^{i} K_{(1)} 
- \frac{12}{3} \phi^5_{(0)}\,\phi_{(1)}\dd^{i} K_{(0)}
- 8\,\pi \, j^{i}_{(1)} &=& 0
\label{eq:moml}
\eea 
where $R$, $R_{ij}$ and $\nabla_i$ refer to the background. In writing 
Eqs.~(\ref{eq:haml}) and (\ref{eq:moml}), we
have used  the fact that the zeroth-order quantities satisfy the constraints.
Notice that, as in the non-linear case, the constraints remain
coupled.

\section{The Initial Data Problem for Relativistic Stellar Perturbations}

We now focus on constructing initial data sets for which the
background is a static and spherically symmetric stellar model, with
4-metric given by
\be
ds^2 = -e^{2\nu}\,dt^2 + e^{2\lambda}\,d\hat r^2 + \hat r^2\,d\theta^2 
+ \hat r^2 \sin^2 \theta\,d \varphi^2,
\label{eq:4metric}
\ee
where the metric coefficients $\nu$ and $\lambda$ are functions 
of the radial coordinate $\hat r$ only. Einstein's equations
for this background reduce to solving three equations. 
The first equation defines the ``mass inside radius $\hat{r}$; 
\be
\frac{dm}{d\hat r} = 4\,\pi\,\hat r^2\,\hat\rho_{(0)} \ .
\label{eq:mass}
\ee
Here $m$ is a function of $\hat r$ that is related to the metric function
$\lambda$ by
\be
e^{-2\lambda} \equiv 1 - \frac{2\,m}{\hat r}.
\ee
Equation~(\ref{eq:mass}) is directly obtained from the Hamiltonian constraint. 
The second equation belongs to 
Einstein's evolution equations and reads
\be
\frac{d\nu}{d\hat r}  = \frac{e^{2\lambda}}{\hat r^2} 
\left(m + 4\,\pi\,\hat r^3\, \hat p_{(0)}\right).
\label{eq:nu}
\ee
Finally, conservation of momentum yields the condition for hydrostatic
equilibrium:
\be
\frac{d\hat p_{(0)}}{d\hat r} = - (\hat \rho_{(0)}+\hat p_{(0)})\,\frac{d\nu}{d\hat r}
\label{eq:press}
\ee
The above stellar structure system of 
equations (the Tolman-Oppenheimer-Volkoff equations)
must be supplemented with an equation
of state. For simplicity, we use the polytropic equation of state
$\hat p = \kappa\,\hat\rho^{\Gamma}$, where $\kappa$ and $\Gamma$ are
the adiabatic constant and index, respectively.
The adiabatic
index $\Gamma$ is related to the polytropic index $n$
by $\Gamma = 1+1/n$. In the specific example provided later, we use $\Gamma =
2$ ($n = 1$).

We now assume that the zero-order quantities ($\hat \rho_{(0)},\,
\hat p_{(0)},\, \lambda,\, \nu$) have been obtained from
solving Eqs.~(\ref{eq:mass}), (\ref{eq:nu}) and (\ref{eq:press}). 
The next step is to solve for the first-order perturbations
from the linearized constraints
(\ref{eq:haml}) and (\ref{eq:moml}).
To facilitate this task, it is convenient to perform
a coordinate transformation and bring the 3-metric in (\ref{eq:4metric})
into the isotropic, conformally flat form:
\be
ds^2 = \phi^4_{(0)} ( dr^2 + r^2\,d\theta^2 + r^2 \sin^2 \theta\,d \varphi^2),
\label{eq:cmetric}
\ee
where the conformal factor is given by
\be
\phi_{(0)} = \left(\frac{\hat r}{r} \right)^{1/2}
\label{eq:cfac}
\ee
and the transformation of the radial-coordinate is obtained from 
\be
\frac{dr}{d\hat r} = e^{\lambda}\, \frac{r}{\hat r}.
\label{eq:drdr}
\ee

The static nature of the background spacetime and the gauge choice 
of a vanishing shift vector imply $K^{(0)}_{ij} =
j^i_{(0)} = 0$ for this background spacetime. 
This leads to considerable simplifications in the following, but
the procedure we describe for 
constructing perturbative initial data can easily be extended 
to include also time-dependent,
spherically symmetric background spacetimes. 

We introduce here our {\em third}
assumption, which is that $A^{ij}_{*(1)} = K_{(1)} = 0$. The vanishing
of the transverse-traceless part of the perturbation to the extrinsic
curvature has direct implications to the gravitational radiation
content of the initial data. 
In a way this assumption can be viewed as ``minimizing'' the amount
of gravitational waves in the initial spacetime. That this may not be 
desirable is obvious, but in order to be able to assign this part of
the free data physically correct values we need a detailed knowledge
of the past history of the system. Such information requires long term,
nonlinear evolutions and is far beyond our present capabilities. The present
assumption is convenient in that it simplifies the calculations considerably. 
Furthermore, we are unlikely to overestimate the amount of gravitational
radiation emerging from true physical systems if we base our estimates
on the present approach. 
The vanishing of the trace
of the extrinsic curvature to first order is less restrictive
physically. It simply implies (using also the fact that $K_{(0)} =
0$) that the slicing of the perturbed spacetime is maximal to first
order. With those further assumptions, the linearized constraints (\ref{eq:haml}) and
(\ref{eq:moml}) decouple and take the form
\bea
\dd^j \dd_{j} \phi_{(1)} & = &  
6\,\pi\, \sigma_{(0)} \,\phi^{-4}_{(0)} \phi_{(1)} 
-2\,\pi\,\phi^{-3}_{(0)}\sigma_{(1)} 
\label{eq:haml2} \\
\dd^i \dd_{i} U &= &\nabla_i V^i \label{eq:u}\\
\dd^j \dd_{j} V^i & = & 8\,\pi\,j^i_{(1)} \label{eq:v}.
\label{eq:moml2}
\eea
where we have decomposed the vector $W^i_{(1)}$ in Eq.~(\ref{eq:moml})
following \cite{bowen,bowen-york} as
\be
W^i_{(1)} = V^i - \frac{1}{4}\nabla^iU .
\label{eq:wsplit}
\ee
From now on, we will drop the label ${}_{(1)}$ in $j^i_{(1)}$ and $W^i_{(1)}$
since the zero-order values for these quantities vanish. 

The set of equations~(\ref{eq:haml2}-\ref{eq:moml2}) constitute a
coupled set of elliptic equations in three dimensions, expressed e.g.,
in the coordinates $(r,\theta,\varphi)$ of the background space. Due to
the spherically symmetric nature of the background, these linearized
equations allow for a separation of variables.  Specifically, we can 
apply a spherical harmonic decomposition of form:
\bea
\phi_{(1)}(r,\theta,\varphi) &=& \sum_{lm} \phi_{(1)}(r)\,Y_{lm}(\theta,\varphi)\\
\sigma_{(1)}(r,\theta,\varphi) &=& \sum_{lm} \sigma_{(1)}(r)\,Y_{lm}(\theta,\varphi)\\
U(r,\theta,\varphi) & = & \sum_{lm} U(r)\,Y_{lm}(\theta,\varphi) \label{eq:uscalar}\\
V^i(r,\theta,\varphi) & = & \sum_{lm} V_1(r)\,e^i_1(\theta,\varphi)
+ r\,V_2(r)\,e^i_2(\theta,\varphi) 
+ r\,V_3(r)\,e^i_3(\theta,\varphi)
\label{eq:vvec} \\
W^i(r,\theta,\varphi) & = & \sum_{lm} W_1(r)\,e^i_1(\theta,\varphi)
+ r\,W_2(r)\,e^i_2(\theta,\varphi) 
+ r\,W_3(r)\,e^i_3(\theta,\varphi)
\label{eq:wvec} \\
j^i(r,\theta,\varphi) & = & \sum_{lm} J_1(r)\,e^i_1(\theta,\varphi) 
+ r\,J_2(r)\,e^i_2(\theta,\varphi),
+ r\,J_3(r)\,e^i_3(\theta,\varphi),
\label{eq:jvec}
\eea
where
\bea
e^i_1 & = & (Y_{lm},\, 0,\, 0) \nonumber \\
e^i_2 & = & \left(0,\, \frac{1}{r^2}\,\partial_\theta Y_{lm},\,
\frac{1}{r^2\,\sin^2\theta}\,\partial_\varphi Y_{lm}\right) \\ 
e^i_3 & = & \left(0,\, -\frac{1}{r^2\,\sin\theta}\,\partial_\varphi Y_{lm},\,
\frac{1}{r^2\,\sin\theta}\,\partial_\theta Y_{lm}\right). \nonumber
\eea
Here $e^i_1$ and $e^i_2$ are the basis vectors of even-parity 
perturbations and $e^i_3$  is the basis for odd-parity perturbations.
In the above expressions and what follows, it is understood that the radial
functions are for a given $(l,m)$. These indices have been 
suppressed for economy in notation.
We note that in terms of the radial functions, Eq.~(\ref{eq:wsplit}) reduces to
\bea
W_1 & = & V_1 - \frac{1}{4}\,\frac{d}{dr}U \nonumber\\
W_2 & = & V_2 - \frac{1}{4}\,U \label{eq:ww}\\
W_3 & = & V_3  \nonumber.
\eea 

After separation of variables, the system of equations~(\ref{eq:haml2}-\ref{eq:moml2}) 
is rewritten as a system of coupled radial elliptic equations:
\bea
\frac{1}{r^2}\frac{d}{dr}\left(r^2\frac{d}{dr}\phi_{(1)}\right) 
- \left[\frac{\l(\l+1)}{r^2} 
+ 6\,\pi\, \sigma_{(0)} \,\phi^{-4}_{(0)}\right]\,\phi_{(1)} 
& = & -2\,\pi\,\phi^{-3}_{(0)}\,\sigma_{(1)} 
\label{eq:hamr} \\
\frac{1}{r^2}\frac{d}{dr}\left(r^2\frac{d}{dr}U\right)
- \frac{\l(\l+1)}{r^2}\,U & = & 
\frac{1}{r^2}\frac{d}{dr}\left(r^2\,V_1\right)
- \frac{\l(\l+1)}{r}\,V_2 \label{eq:uu}\\
\frac{1}{r^2}\frac{d}{dr}\left(r^2\frac{d}{dr}V_1\right)
- \frac{[\l(\l+1)+2]}{r^2}\,V_1 + 2\,\frac{\l(\l+1)}{r^2}\,V_2 & = & 
8\,\pi\,J_1 \label{eq:v1} \\
\frac{1}{r^2}\frac{d}{dr}\left(r^2\frac{d}{dr}V_2\right)
- \frac{\l(\l+1)}{r^2}\,V_2 + \frac{2}{r^2}\,V_1 & = & 
8\,\pi\,J_2 \label{eq:v2}\\
\frac{1}{r^2}\frac{d}{dr}\left(r^2\frac{d}{dr}V_3\right)
- \frac{\l(\l+1)}{r^2}\,V_3 & = & 
8\,\pi\,J_3 \label{eq:v3}.
\eea

Equations~(\ref{eq:hamr}-\ref{eq:v3}) fully characterize, for each
$(l,m)$ harmonic, initial data $(\phi_{(1)},U,V_{i})$ to first perturbative
order, once the background conformal factor $\phi_{(0)}$ and density
$\sigma_{(0)}$ as well as the fluid perturbations $\sigma_{(1)}$ and $J_{i}$ are
specified.
Outside the sources, the solutions to the above
equations are $U = u\,r^{a}$,
$V_1 = v_1\,r^{b}$, $V_2 = v_2\,r^{c}$ and
$V_3 = v_3\,r^{d}$ with
$u$, $v_1$, $v_2$ and $v_3$ constants.
In order to have regular solutions
for $r \rightarrow \infty$, one needs $a = d =  -(\l+1)$,
$b = c = -l$ and
$v_1 = \l\,v_2$, where we assume that $\l \ge 1$.
As we shall later see, 
the corresponding interior
solutions for head-on and inspiral close-limit collisions
exhibit the same scaling; that is,
$V_1 = \l\,V_2$ as well as $J_1 = \l\,J_2$.
With this assumption, the system of equations
(\ref{eq:uu}-\ref{eq:v3}) reduces to
\bea
\frac{1}{r^2}\frac{d}{dr}\left(r^2\frac{d}{dr}U\right)
- \frac{\l(\l+1)}{r^2}\,U & = &
\frac{d}{dr}V_1
- \frac{(\l-1)}{r}\,V_1 \label{eq:uu2}\\
\frac{1}{r^2}\frac{d}{dr}\left(r^2\frac{d}{dr}V_1\right)
- \frac{\l(\l-1)}{r^2}\,V_1 & = &
8\,\pi\,J_1 \label{eq:vv2}\\
\frac{1}{r^2}\frac{d}{dr}\left(r^2\frac{d}{dr}V_3\right)
- \frac{\l(\l+1)}{r^2}\,V_3 & = &
8\,\pi\,J_3 \label{eq:vv3}.
\eea

\section{Correspondence with Regge-Wheeler variables}
\label{sec:regge}

Before we proceed to present examples of initial data sets constructed 
using the above approach, we want to
establish the correspondence between our variables and the 
standard Regge-Wheeler variables. This is relevant since the
Regge-Wheeler notation (and the associated gauge) is customarily used in
perturbative evolutions for spherical stellar models \cite{allen}. To this end,
consider a spatial tensor $T_{ij}$ such that 
$T_{ij}(\hat r,\,\theta,\,\varphi) = 
T_{ij}^{(0)}(\hat r) + T_{ij}^{(1)}(\hat r,\,\theta,\,\varphi)$. The  
perturbations of this tensor can be decomposed as
\be
T^{(1)}_{ij} = t_1(\hat r)\,f^1_{ij} + \hat r\,t_2(\hat r)\,f^2_{ij}
+ \hat r^2\,t_3(\hat r)\,f^3_{ij} + \hat r^2\,t_4(\hat r)\,f^4_{ij}
+ t_5(\hat r)\,f^5_{ij} + \hat t_6(\hat r)\,f^6_{ij},
\label{eq:even}
\ee
where
\be
 \mbox{$f^1_{ij}$} = \left(\matrix{
                    Y_{lm} & 0 & 0\cr
                    0  &   0   &  0 \cr
                    0  &   0   &  0 \cr
                }\right),
\label{eq:f1}
\ee

\be
 \mbox{$f^2_{ij}$} = \left(\matrix{
                    0            & \partial_\theta Y_{lm} &  \partial_\varphi Y_{lm} \cr
                    \hbox{symm}  &   0   &  0 \cr
                    \hbox{symm}  &   0   &  0 \cr
                }\right),
\label{eq:f2}
\ee

\be
 \mbox{$f^3_{ij}$} = \left(\matrix{
                    0  & 0      & 0\cr
                    0  & Y_{lm} & 0 \cr
                    0  & 0      & \sin^2\theta\,Y_{lm} \cr
                }\right),
\label{eq:f3}
\ee

\be
 \mbox{$f^4_{ij}$} = \left(\matrix{
                    0  & 0      & 0\cr
                    0  & \partial^2_\theta\,Y_{lm} & (\partial_\theta\partial_\varphi
                         - \cot\theta\partial_\varphi)\,Y_{lm} \cr
                    0  & \hbox{symm} & (\partial^2_\varphi + \sin\theta\,\cos\theta\,
                         \partial_\theta)\,Y_{lm} \cr
                }\right)
\label{eq:f4}
\ee

\be
 \mbox{$f^5_{ij}$} = \left(\matrix{
                    0            & -\frac{1}{\sin\theta}\partial_\varphi Y_{lm} 
                                 &  \sin\theta\partial_\theta Y_{lm} \cr
                    \hbox{symm}  &   0   &  0 \cr
                    \hbox{symm}  &   0   &  0 \cr
                }\right),
\label{eq:f5}
\ee

\be
 \mbox{$f^6_{ij}$} = \left(\matrix{
                    0  & 0      & 0\cr
                    0  & \frac{1}{\sin\theta}
                    (\partial_\theta\partial_\varphi-\cot\theta)\,Y_{lm} 
                    & \frac{\sin\theta}{2} \left(\frac{1}{\sin^2\theta}
                      \partial^2_\varphi + \cot\theta\partial_\theta
                       -\partial^2_\theta \right)\,Y_{lm} \cr
                    0  & \hbox{symm} & -\sin\theta(\partial_\theta\partial_\varphi 
                      - \cot\theta \partial_\varphi)\,Y_{lm} \cr
                }\right).
\label{eq:f6}
\ee
Above, $f^1_{ij}$, $f^2_{ij}$, $f^3_{ij}$, and $f^4_{ij}$
represent the even-parity tensor spherical harmonics and
$f^5_{ij}$ and $f^6_{ij}$ the odd-parity counterparts.

Using the Regge-Wheeler notation, the perturbations of the 
spatial metric (in physical space) read
\be
\hat h_{ij}^{(1)} = \sum_{lm} e^{2\lambda}\,H_2(\hat r)\,f^1_{ij} 
+ h_1^{even}(\hat r)\,f^2_{ij}
+ \hat r^2\,K(\hat r)\,f^3_{ij} + \hat r^2\,G(\hat r)\,f^4_{ij}
+ h_1^{odd}(\hat r)\,f^5_{ij} + h_2(\hat r)\,f^6_{ij},
\label{eq:rw}
\ee
where $K$ must not be confused with the trace of the extrinsic curvature.
From the previous section, we have that the spatial metric can be constructed from
\bea
\hat h_{ij} & = & \hat h_{ij}^{(0)} + \hat h_{ij}^{(1)} 
             =  \phi^4\, h_{ij} 
             =  (\phi_{(0)} + \phi_{(1)})^4\,\, h_{ij}^{(0)}\nonumber \\
            & = & \phi_{(0)}^4\,h_{ij}^{(0)} + 4\,\phi_{(0)}^3\phi_{(1)}\, h_{ij}^{(0)} \\
            & = & \hat h_{ij}^{(0)} + 4\,\frac{\phi_{(1)}}{\phi_{(0)}}\,\hat h_{ij}^{(0)},
\eea
where $\phi_{(0)} = (\hat r/r)^{1/2}$, $\hat h^{(0)}_{ij} = \hbox{diag}(
e^{2\lambda},\,\hat r^2,\, \hat r^2\,\sin^2\theta)$ and 
$\phi_{(1)} = \sum_{lm} \phi_{(1)}(r)\,Y_{lm}$ with $\phi_{(1)}(r)$
a solution of the radial equation (\ref{eq:hamr}). Thus, our approach to construct
initial data yields spatial metric perturbations of the form
\be
\hat h_{ij}^{(1)} = \sum_{lm} 4\,\phi_{(1)}(r)\left(\frac{\hat r}{r}\right)^{1/2}
(e^{2\lambda}\,f^1_{ij} + \hat r^2\,f^3_{ij}).
\label{eq:hrw}
\ee
Comparison of the metric perturbations (\ref{eq:hrw}) with (\ref{eq:rw})
shows that our procedure for constructing initial data yields 
in terms of the Regge-Wheeler notation 
$h_1^{odd} = h_1^{even} = h_2 = G = 0$ and
\be
H_2 = K = 4\,\phi_{(1)}(r)\left(\frac{\hat r}{r}\right)^{1/2}.
\ee

Consider now the extrinsic curvature. As mentioned before,
the extrinsic curvature vanishes to zero-order. 
In addition, we made the choice of
having vanishing first-order
trace $K_{(1)}$ and transverse-traceless parts $A^{ij}_{*(1)}$. Thus,
the extrinsic curvature is completely determined by the vector $W^i$ 
and the conformal factor of the background space $\phi_{(0)}$ from
\be
\hat K_{ij} = \phi^{-2}_{(0)}\,(\l W)_{ij}.
\ee
In terms of tensor spherical harmonics, the extrinsic curvature reads
\be
\hat K_{ij} = \sum_{lm} \left(\frac{r}{\hat r}\right) \left[
       k_1(r)\,f^1_{ij} 
  + r\,k_2(r)\,f^2_{ij}
+ r^2\,k_3(r)\,f^3_{ij} 
+ r^2\,k_4(r)\,f^4_{ij}
      +k_5(r)\,f^5_{ij} 
+      k_6(r)\,f^6_{ij}
\right],
\label{eq:krw}
\ee
where
\bea
k_1 & = & \frac{2}{3\,r}\left[2\,r\,\frac{d}{dr} W_1 
          + \l(\l+1)\,W_2 - 2\,W_1\right] \nonumber \\
k_2 & = & \frac{1}{r}\left[r\,\frac{d}{dr}W_2 
          + W_1 - W_2\right] \nonumber \\
k_3 & = & \frac{2}{3\,r}\left[-r\,\frac{d}{dr}W_1 
          + W_1 + l(l+1)\,W_2\right] \\
k_4 & = & \frac{2}{r}\,W_2 \nonumber \\
k_5 & = & r\,\frac{d}{dr}W_3 - W_3 \nonumber \\
k_6 & = & -2\,r\,W_3\nonumber . \\
\eea

Before proceeding, it is appropriate to discuss 
the implications of the assumptions we imposed 
for the presence of even and odd parity perturbations
in the initial data.

Consider first the perturbations of the 3-metric: The assumption of
vanishing perturbations of the conformal 3-metric implies (as seen
from Eq.~(\ref{eq:hrw})), that $h_1^{odd} = h_1^{even} = h_2 = G =
0$. This means that all odd parity perturbations of the 3-metric must 
vanish on the initial surface. This does not however exclude the 
presence of odd-parity
perturbations. Such perturbations may enter via the Regge-Wheeler variable
$h_0$, which is freely specifiable. 
Similarly for even-parity perturbations, the Regge-Wheeler
quantities $H_0$ and $H_1$ are not part of
the constraints and can be chosen freely. A choice of 
these three variables correspond to choosing a slicing for the spacetime, i.e.
specifying the lapse and the shift vector. Specifically, Regge-Wheeler
gauge corresponds to $h_1^{even}=G=h_2=0$ above, as well as perturbed lapse
\be
\delta \alpha = e^{2\nu}H_0 \ ,
\ee
and shift vector
\be
\delta \beta^i = H_1 e_1^i + h_0 e_3^i \ .
\ee 

Furthermore, there are no constraints on the odd or even parity
nature of the extrinsic curvature initial data. This follows immediately from
Eq.~(\ref{eq:krw}) since the coefficients $k_1$ through $k_6$ are in
general non-vanishing. As a consequence, even if the initial 3-metric
has vanishing odd-parity perturbations, their time evolution will in
general include such perturbations.

\section{Sample Initial data sets: Colliding neutron stars}

We will now apply the method for constructing 
initial data, that was presented in
Sec.~III, to a case of astrophysical relevance. 
We consider collisions of neutron stars
under the close-limit approximation.
Related results have already been discussed
by Allen et al \cite{nsmergers}. That study was specialized to the 
case of head-on collisions. Here, we present a general discussion of 
neutron-star close-limit initial data, and give explicit results 
for both boosted head-on and inspiralling collisions.

As stated at the end of Sec.~III, initial data are obtained
by solving the system of equations 
equations~(\ref{eq:hamr}-\ref{eq:v3}).
The input for these equations are 
the background conformal factor $\phi_{(0)}$ and density
$\sigma_{(0)}$ and, in addition, the fluid perturbations $\sigma_{(1)}$ and $J_{i}$.
For close-limit collisions, one specifies the background 
from the outcome of the collisions. 
The fluid perturbations
$\sigma_{(1)}$ and $J_{i}$, on the other hand, are obtained by ``subtracting" 
the background from a suitable superposition of stars that represent
the initial configuration. A suitable way of relating the two initial stars to the final configuration
was presented in \cite{nsmergers}, and we refer the reader to that paper
for further details.  

Let us first consider the perturbation to the background density.
Neglecting complicating factors, such as the effects from tidal deformations,
we approximate the total density of the binary
system with the following superposition of density profiles
of isolated neutron stars \cite{nsmergers}:
\be
\sigma( r^i) = \sigma_*( r^i-\xi^i)+\sigma_*( r^i+\xi^i)
-\left[\sigma_*( r^i-\xi^i)\,\sigma_*( r^i+\xi^i)\right]^{1/2}\,,
\label{eq:rhotot}
\ee
where $\sigma_*$ is the conformally transformed density profile of the
colliding neutron stars in isolation located 
a distant $\xi^i$ in conformal space.
This functional form 
(\ref{eq:rhotot}) for the total
density is chosen since it leads to 
the correct zero-separation and infinite-separation
limits. That is, for zero-separation
$\sigma \rightarrow \sigma_*$, and for large separations
$\sigma(r^i) \rightarrow \sigma_*( r^i-\xi^i)+\sigma_*( r^i+\xi^i)$.
We now introduce the close-limit approximation, $\xi^i \ll 1$, 
and write
\be
\sigma_*( r^i\pm\xi^i) = \sigma_*( r^i) \pm  \xi^i\,\nabla_i\,\sigma_*(
 r^i)
         +\frac{1}{2}\,\xi^i\,\xi^j\,\nabla_i\,
         \nabla_j\,\sigma_*( r^i)\,.
\label{eq:rhotaylor}
\ee
Therefore, Eq.~(\ref{eq:rhotot}) takes the form
\be
\sigma =  \sigma_*
+ \frac{1}{2}\,(\xi^i\,\nabla_i\,\sigma_*)^2
+ \frac{1}{2}\,\xi^i\,\xi^j\,\nabla_i\,\nabla_j\,\sigma_* \,.
\ee
Given this total energy, the density perturbation is obtained from
\bea
\sigma_{(1)} & = &\sigma-\sigma_{(0)} \nonumber \\
& = &   \sigma_* - \sigma_{(0)}
+ \frac{1}{2}\,(\xi^i\,\nabla_i\,\sigma_*)^2
+ \frac{1}{2}\,\xi^i\,\xi^j\,\nabla_i\,\nabla_j\,\sigma_*\,.
\label{eq:rho11}
\eea
Similarly, the momentum density is given by the superposition of 
momentum densities
of boosted isolated stars:
\be
j^i(r^k) = j^i_+(r^k+\xi^k) + j^i_-(r^k-\xi^k) \,.
\label{eq:jtot}
\ee
In this case, there is no need for a counterpart of the last term in (\ref{eq:rhotot}).
For both, head-on and inspiral collisions,
$j^i_+  = - j^i_- = j^i_*$; thus,
\be
j^i(r^k) = j^i_*(r^k+\xi^k) - j^i_*(r^k-\xi^k) \,,
\label{eq:jtot2}
\ee
which obviously has the 
appropriate zero-separation limit, namely $j^i(r^k) \rightarrow 0$ as
$\xi^i \rightarrow 0$.
Once again, we apply the close-limit condition and approximate
\be
j^i_*( r^k\pm\xi^k) = j^i_{*}( r^k) \pm  \xi^j\,\nabla_j\,j^i_{*}(r^k)
         +\frac{1}{2}\,\xi^l\,\xi^m\,\nabla_l\,\nabla_m\,j^i_{*}( r^k)\,.
\ee
Therefore, 
\be
j^i = 2\,\xi^j\,\nabla_j\,j^i_{*}\,.
\label{eq:j11}
\ee
Notice that Eq.~(\ref{eq:j11}) directly gives the momentum density perturbation
because, by construction, the background is static.
 
\subsection{Head-on Collision Initial Data}

For simplicity, we assume that the collision takes place along
the $z$-axis. Therefore,
\bea
\xi^i &=& \xi\,z^i \label{eq:vecxi} \\
j^i_* &=& -J_*\,z^i \label{eq:vecj}\,,
\eea
where
\be
z^i = \left(\cos{\theta},\,-\frac{1}{ r}\,\sin{\theta},\, 0\right)\, 
\label{eq:vecz}
\ee
is a unit vector along the $z$-axis.

Substitution of (\ref{eq:vecxi}) into (\ref{eq:rho11}) yields
\bea
\sigma_{(1)} &=& \sigma_* - \sigma_{(0)}
+ \frac{1}{2}\,\xi^2\,\left[\cos^2{\theta}\,\left(\frac{d}{d
r}\sigma_*\right)^2 +\cos^2{\theta}\,\frac{d^2}{d
r^2}\sigma_* +\sin^2{\theta}\,\frac{1}{ r}\frac{d}{d
r}\sigma_*\right] \,. \nonumber \\
&=& \sqrt{4\,\pi}\,Y_{00}\,
\bigg[ \sigma_* - \sigma_{(0)}
+ \frac{1}{6}\,\xi^2\,\biggl\lbrace\frac{d^2}{d r^2}\sigma_*
+ \left(\frac{d}{d r}\sigma_*\right)^2
+ \frac{2}{ r}\frac{d}{d r}\sigma_*\biggr\rbrace\bigg]
 \nonumber \\
&+& \frac{2}{15}\,\sqrt{5\,\pi}\,Y_{20}\,\xi^2\,
\left[\frac{d^2}{d r^2}\sigma_*
+ \left(\frac{d}{d r}\sigma_*\right)^2
-\frac{1}{ r}\frac{d}{d r}\sigma_*\right] \, .
\label{eq:rho_b}
\eea
It is clear from (\ref{eq:rho_b}) that
the conformal density perturbation has 
a monopole contribution ($m=0,\, \l= 0$). Since our main interest
is associated with the gravitational waves generated during the
merger, we will ignore this contribution;  
it  obviously does not lead to any gravitational waves.
The remaining part 
is the radiative quadrupole ($m=0,\, \l= 2$) perturbation:
\be
\sigma_{(1)} =
\frac{2}{15}\,\sqrt{5\,\pi}\,\xi^2\,
\left[\frac{d^2}{d r^2}\sigma_*
+ \left(\frac{d}{d r}\sigma_*\right)^2
-\frac{1}{ r}\frac{d}{d r}\sigma_*\right]\,.
\label{eq:rho_final}
\ee
This  initial data set is discussed in considerable detail
in \cite{nsmergers}.

Let us now consider the momentum density perturbation. Substitution of
(\ref{eq:vecj}) into (\ref{eq:j11}) yields
\bea
j^i &=& -2\,\xi\,z^j\,\nabla_j(\,J_*\,z^i)\nonumber \\
    &=& -2\,\xi\,z^i\,z^j\,\nabla_j\,J_* \nonumber \\
    &=& -2\,\xi\,\frac{d}{dr}J_*\,\left(\cos^2{\theta},
-\frac{1}{r}\sin{\theta}\,\cos{\theta},0 \right)\nonumber \\
&=& 
-\frac{4}{3}\,\sqrt{\pi}\,\xi\,\frac{d}{dr}J_*\,\left(Y_{00},0,0\right) 
-\frac{8}{15}\,\sqrt{5\,\pi}\,\xi\,\frac{d}{dr}J_*\,\left(
Y_{20}, \frac{2}{r}\partial_\theta Y_{20}, 0\right)\,.
\label{eq:jhead}
\eea
As expected and in agreement with the density perturbation, 
we again  have a monopole momentum density perturbation that we shall ignore.
Comparing the quadrupole terms in (\ref{eq:jhead}) with (\ref{eq:jvec}), we 
have 
$J_3 = 0$ and 
\be
J_1 = 2\,J_2 = -\frac{8}{15}\,\sqrt{5\,\pi}\,\xi\,\frac{d}{dr}J_*\,.
\label{eq:jhfinal}
\ee

From Eq.~(\ref{eq:sigma}) and (\ref{eq:j}) and the conformal transformations
(\ref{eq:csigma}) and (\ref{eq:cj}), the expressions for
$\sigma_*$ and $J_*$ 
in Eqs.~(\ref{eq:rho_final}) and (\ref{eq:jhfinal})  
are given by:
\bea
\sigma_* &=& \hat\rho_{*}\,\phi^{8}_{(0)}
\label{eq:sigmamag}\\
J_* &=& (\hat\rho_{*} + \hat p_{*})\,v\,\phi^{10}_{(0)}\,,
\label{eq:jmag}
\eea
with $v$ the magnitude of the collision velocity. 
In writing the above expressions, we used that
to zero-order $\hat u^i = 0$ and $\hat\gamma = 1 + O(\hat u^2)$.
Notice that the conformal factor in connection with the background star, 
$\phi_{(0)}$, was the one used in
Eqs.~(\ref{eq:sigmamag}) and (\ref{eq:jmag}) to transform the physical TOV solutions
of the colliding stars since it is the background star that provides the 
conformal space where our calculations are performed.
Also important is to notice that, once the background and colliding neutron stars models 
have been completely determined, there are only two parameters that
characterize the initial data: the separation $\xi$ and the velocity $v$.

\subsection{Inspiral Collision Initial Data}

For this case, we assume that the initial configuration is such that
the neutron stars are along the $x$-axis and their momentum pointing
along the $y$-axis. Then
\bea
\xi^i &=& \xi\,x^i \label{eq:vecxi_r} \\
j^i_* &=& J_*\,y^i \label{eq:vecj_r}\,,
\eea
where
\bea
x^i &=& \left(\sin{\theta}\,\cos{\varphi},
\frac{1}{r}\cos{\theta}\,\cos{\varphi},
-\frac{\sin{\varphi}}{r\,\sin{\theta}}\right)
\label{eq:vecx} \\
y^i &=& \left(\sin{\theta}\,\sin{\varphi},
\frac{1}{r}\cos{\theta}\,\sin{\varphi},
\frac{\cos{\varphi}}{r\,\sin{\theta}}\right)\,.
\label{eq:vecy}
\eea
are unit vectors along the $x$-axis and $y$-axis, respectively.
Substitution of (\ref{eq:vecxi_r}) and (\ref{eq:vecj_r}) into
(\ref{eq:rho11}) yields
\bea
\sigma_{(1)} &=& \sigma_* - \sigma_{(0)}
+ \frac{1}{2}\,\xi^2\,\bigg[\sin^2{\theta}\,\cos^2{\varphi}\,\biggl\lbrace\left(\frac{d}{d
r}\sigma_*\right)^2 +\frac{d^2}{d r^2}\sigma_*\biggr\rbrace \nonumber \\
&+&(\cos^2{\theta}\,\cos^2{\varphi}+\sin^2{\varphi})\frac{1}{ r}\frac{d}{d
r}\sigma_*\bigg] \nonumber \\
&=& \sqrt{4\,\pi}\,Y_{00}\,
\bigg[ \sigma_* - \sigma_{(0)}
+ \frac{1}{6}\,\xi^2\,\biggl\lbrace\frac{d^2}{d r^2}\sigma_*
+ \left(\frac{d}{d r}\sigma_*\right)^2
+ \frac{2}{ r}\frac{d}{d r}\sigma_*\biggr\rbrace\bigg] 
 \nonumber \\
&+& \left(-\sqrt{\frac{\pi}{5}}\,Y_{20}+
\sqrt{\frac{2\,\pi}{15}}\hbox{Re}Y_{22}\right)\,\xi^2\,
\left[\frac{d^2}{d r^2}\sigma_*
+ \left(\frac{d}{d r}\sigma_*\right)^2
-\frac{1}{ r}\frac{d}{d r}\sigma_*\right] \, .
\label{eq:rho_i}
\eea
As with the head-on collisions, we concentrate on the
radiative quadrupole ($m=2,\,l=2$) term, so the density perturbation
$\sigma_{(1)}$ source of Eq.~(\ref{eq:hamr}) is given by:
\be
\sigma_{(1)} = \sqrt{\frac{2\,\pi}{15}}\,\xi^2\,
\left[\frac{d^2}{d r^2}\sigma_*
+ \left(\frac{d}{d r}\sigma_*\right)^2
-\frac{1}{ r}\frac{d}{d r}\sigma_*\right]\,.
\label{eq:rho_ins}
\ee
Notice that the only difference between (\ref{eq:rho_ins}) and the corresponding source term
in the head-on collision case, i.e. Eq.~(\ref{eq:rho_final}), is a numerical factor.

For the momentum density perturbation, substitution of
(\ref{eq:vecj_r}) into (\ref{eq:j11}) yields
\bea
j^i &=& 2\,\xi\,x^j\,\nabla_j(\,J_*\,y^i)\nonumber \\
    &=& 2\,\xi\,y^i\,x^j\,\nabla_j\,J_* \nonumber \\
    &=& 2\,\xi\,\frac{d}{dr}J_*\,\left(\sin^2{\theta}\,\sin{\varphi}\,\cos{\varphi},
\frac{1}{r}\sin{\theta}\,\cos{\theta}\,\sin{\varphi}\,\cos{\varphi},
\frac{1}{r}\cos^2{\varphi}\right)\nonumber \\
&=& -\sqrt{\frac{4\,\pi}{3}}\,\xi\,\frac{d}{dr}J_*\,
\left(0,0,\frac{1}{r\,\sin{\theta}}\partial_\theta\,Y_{10}\right)
+ 4\sqrt{\frac{2\,\pi}{15}}\,\xi\,\frac{d}{dr}J_*\,\hbox{Im}\left(Y_{22},
\frac{1}{r}\partial_\theta\,Y_{22},
\frac{1}{r\,\sin^2{\theta}}
\partial_\varphi\,Y_{22}\right)\,.
\label{eq:jrot1}
\eea
Comparing (\ref{eq:jrot1}) with (\ref{eq:jvec}), we deduce 
for the quadrupole term ($m=2,\,l=2$)
$J_3 = 0$ and 
\be
J_1 = 2\,J_2 = 4\sqrt{\frac{2\,\pi}{15}}\,\xi\,\frac{d}{dr}J_*\,,
\label{eq:jquadra}
\ee
The dipole term ($m=0,\,l=1$)
does not contribute to the emerging gravitational radiation
and can be ignored. Once again, the momentum density perturbation for the
inspiral case only differs from the head-on case by a numerical factor. 
The only non-trivial differences in the initial data will then
arise from the $Y_{lm}$'s since in one case 
$m=0$ (head-on) and for the other $m=2$ (inspiral).
The quantities $\sigma_*$ and $J_*$ in (\ref{eq:rho_ins}) and
(\ref{eq:jquadra}) are obtained as in the
head-on collision case, namely from Eqs.~(\ref{eq:sigmamag}) and (\ref{eq:jmag})
respectively.

\begin{figure}[fig1]
\leavevmode
\\
\epsfxsize=0.8\textwidth
\epsfbox{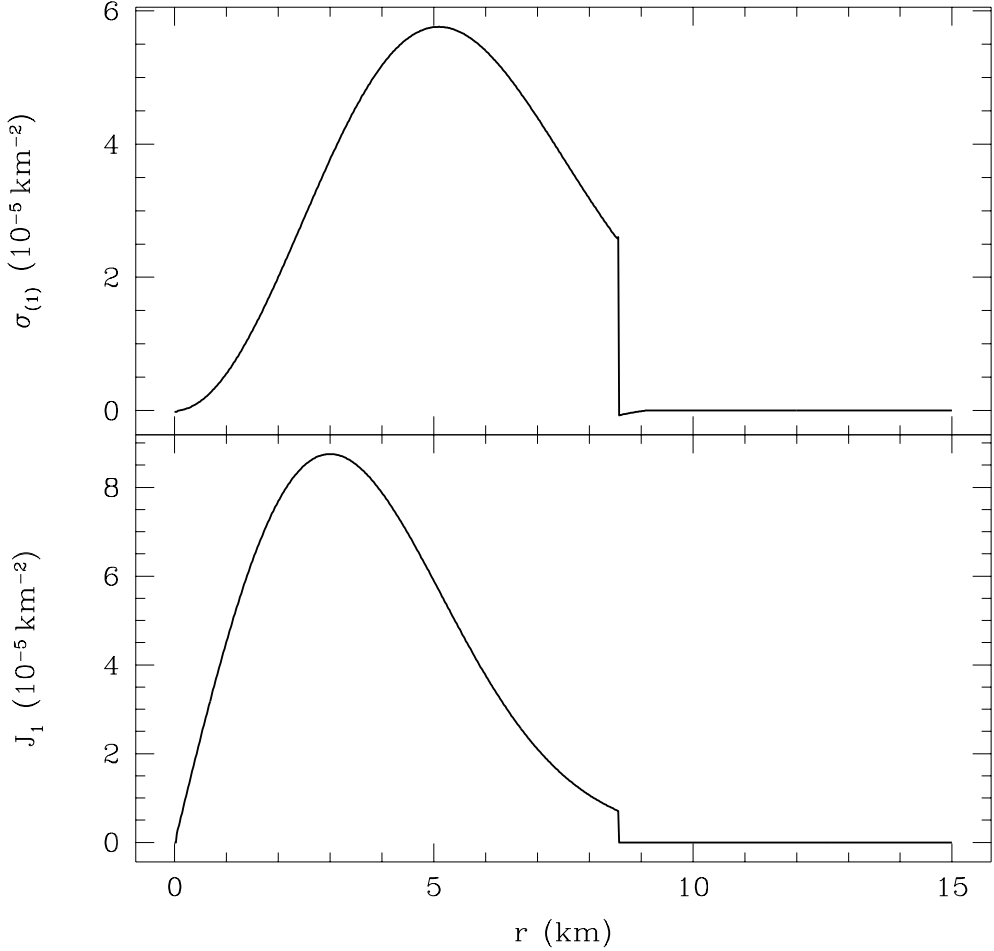}
\caption[figure1]{\label{fig1} 
Density $\sigma_{(1)}$ and 
momentum density $J_1$ perturbations 
(see Eqs.~(\ref{eq:rho_final} and \ref{eq:jhfinal} in the text) for
the close-limit, boosted, head-on collision of neutron stars
with initial separation $\xi = 0.1\,R_{(0)}$ and velocity $v = 0.1\,c$.
The corresponding perturbations for the inspiral case differ from these 
quantities by constant numerical factors.
The TOV parameters for the
background and colliding stars are:
$\rho^{(0)}_c = 2.69\times 10^{15}\,\hbox{g/cm}^3$ and  $\kappa_{(0)} = 100\,\hbox{km}^2 $.
For these parameters, the mass and radius of the background star are
$M_{(0)} = 1.24 \,M_\odot$ and $R_{(0)} = 9.0\,\hbox{km}$, respectively. The initial
colliding stars, which are displaced a distance $0.1R_0$ from the
center of mass, follow from $\rho^*_c = 2.98\times 10^{15}\,\hbox{g/cm}^3$
and  $K_* = 90.25\,\hbox{km}^2 $. With these parameters, the colliding stars
have a mass and radius of $M_* = 1.17 \,M_\odot$ and
$R_0* = 8.58\,\hbox{km}$, respectively.
} \end{figure}

\begin{figure}[fig2]
\leavevmode
\\
\epsfxsize=0.8\textwidth
\epsfbox{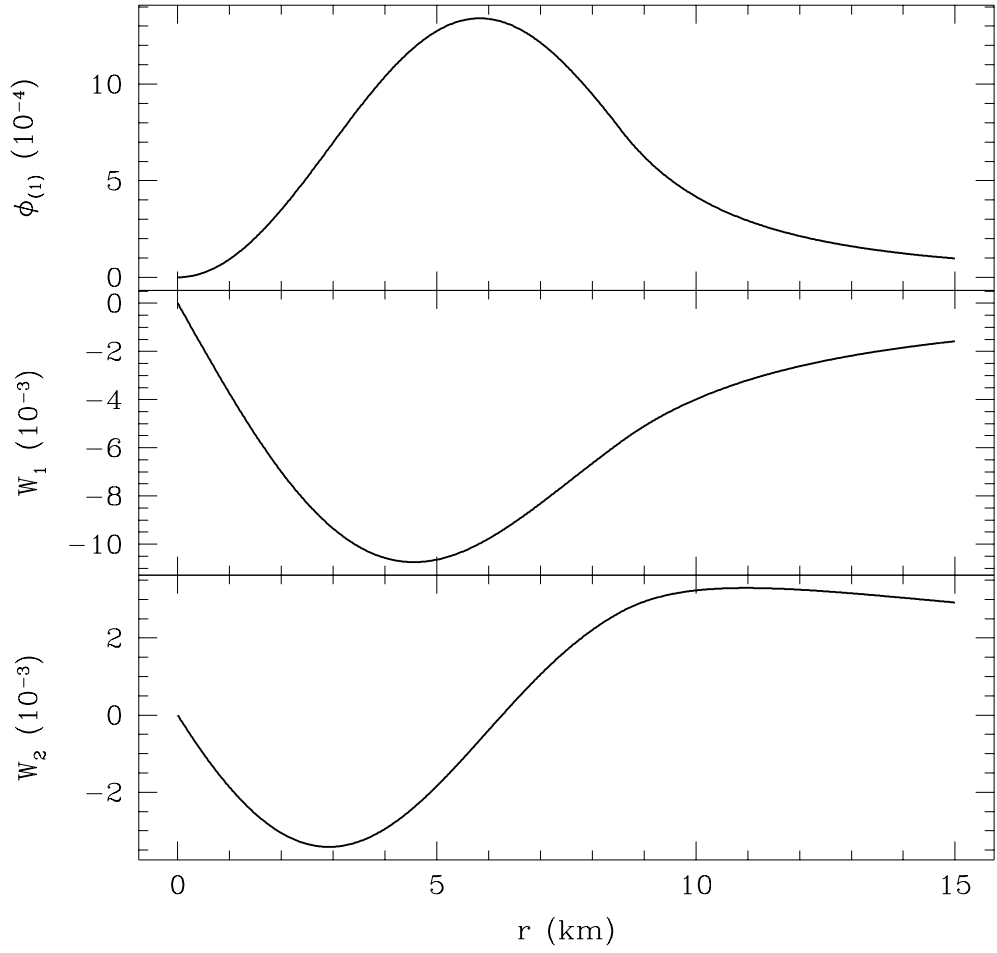}
\caption[figure2]{\label{fig2}
Conformal perturbation  $\phi_{(1)}$ and 
harmonic components $W_1$ and $W_2$ of the 
vector $W^i$ from the solution to the linearized constraints for 
the close-limit collision of neutron stars corresponding to
the perturbations in Fig.\ref{fig1}.
} \end{figure}

Figure~\ref{fig1} shows profiles 
of the density perturbation $\sigma_{(1)}$ 
and the momentum density perturbation $J_1$ for the close-limit, boosted, head-on collision.
Recall that for inspiral and head-on collisions $J_1 = 2\,J_2$ and $J_3 = 0$.
The perturbations $\sigma_{(1)}$ and $J_1$ were calculated from neutron stars
with initial separation $\xi = 0.1\,R_{(0)}$ and velocity $v = 0.1\,c$.
The corresponding perturbations for the inspiral case only differ from 
the perturbations shown in Fig.~\ref{fig1} by a constant numerical factor
(compare Eqs.~(\ref{eq:rho_final}) and (\ref{eq:jhfinal}) with
Eqs.~(\ref{eq:rho_ins}) and (\ref{eq:jquadra}).
The TOV parameters for the
background and colliding stars are:
$\rho^{(0)}_c = 2.69\times 10^{15}\,\hbox{g/cm}^3$ and  $\kappa_{(0)} = 100\,\hbox{km}^2 $.
For these parameters, the mass and radius of the background star are
$M_{(0)} = 1.24 \,M_\odot$ and $R_{(0)} = 9.0\,\hbox{km}$, respectively. The initial
colliding stars, which are displaced a distance $0.1R_0$ from the
center of mass, follow from $\rho^*_c = 2.98\times 10^{15}\,\hbox{g/cm}^3$
and  $K_* = 90.25\,\hbox{km}^2 $. With these parameters, the colliding stars 
have a mass and radius of $M_* = 1.17 \,M_\odot$ and
$R_0* = 8.58\,\hbox{km}$, respectively.
Figure~\ref{fig2} shows the solutions to the
conformal perturbation  $\phi_{(1)}$ and the harmonic components 
$W_1$ and $W_2$ (see Eq.~\ref{eq:ww}) of the 
vector $W^i$ for the close-limit collision of neutron stars corresponding to
the perturbations in Fig.\ref{fig1}.

\section{Concluding remarks}

In this paper, we have presented a framework for constructing 
initial data relevant for perturbative studies of neutron stars. 
Our approach was to ``linearize'' Lichnerowicz-York's standard
procedure for the initial-value problem in General Relativity, 
and it facilitates (to a certain extent) setting astrophysical
initial data for perturbation evolutions, cf. \cite{allen}.  
It is straightforward to compare our method (as well as the results)
to the fully nonlinear one, which is important since a main
motivation for perturbation studies is to provide benchmark tests
for nonlinear numerical relativity. 

As examples of interesting initial data that 
can  be constructed from our equations, we constructed
data for merging neutron stars in the close-limit approximation.
The simplest case of these data sets, that describe
head-on collision of two initially static stars, has already been extensively
discussed
in \cite{nsmergers}. No studies of the more 
general data with initial momentum and for inspiralling 
collisions have yet been performed. Such simulations should
obviously 
be carried out, and we hope  to be able to discuss
the relevant results, as well as possible extensions of the framework 
developed in this paper to, for example, rotating configurations, in the 
near future.

\section*{Acknowledgments}

We thank Johannes Ruoff for his assistance with some of the tensor harmonics
manipulations. This work was partially supported by
NATO grants CRG960260 and CRG971092, as well as NSF grants PHY9800973 and PHY9357219.

\bibliographystyle{unsrt}

\end{document}